\documentclass{IEEEcsmag}

\jvol{27}
\jnum{1}
\jmonth{January/February}
\pubyear{2025}
\jname{IT Professional}

\usepackage[colorlinks,urlcolor=blue,linkcolor=blue,citecolor=blue]{hyperref}
\expandafter\def\expandafter\UrlBreaks\expandafter{\UrlBreaks\do\/\do\*\do\-\do\~\do\'\do\"\do\-}
\usepackage{upmath,color}
\usepackage{graphicx}
\usepackage{multirow}
\let\labelindent\relax
\usepackage{enumitem}
\usepackage{changepage}
\usepackage{listings}
\usepackage[table]{xcolor}
\usepackage{colortbl}

\def\thickhline{\noalign{\hrule height.9pt}}

\begin{document}

\title{Landscape and Taxonomy of Prompt Engineering Patterns in Software Engineering}

\author{Yuya Sasaki}
\affil{Waseda University, Tokyo 169-8555, Japan \\ sasaki.y@ruri.waseda.jp}

\author{Hironori Washizaki}
\affil{Waseda University, Tokyo 169-8555, Japan \\ washizaki@waseda.jp}

\author{Jialong Li}
\affil{Waseda University, Tokyo 169-8555, Japan \\ lijialong@fuji.waseda.jp}

\author{Nobukazu Yoshioka}
\affil{Waseda University, Tokyo 169-8555, Japan \\ nobukazu@engineerable.ai}

\author{Naoyasu Ubayashi}
\affil{Waseda University, Tokyo 169-8555, Japan \\ ubayashi@aoni.waseda.jp}

\author{Yoshiaki Fukazawa}
\affil{Waseda University, Tokyo 169-8555, Japan \\ fukazawa@waseda.jp}

\begin{abstract}
Advancements in large language models (LLMs) have enhanced their ability to handle ambiguous user instructions. However, effective prompt patterns remain crucial for usability and comprehension. This article presents a taxonomy of prompt engineering patterns for software engineering. It is based on a systematic literature review that was conducted in early 2023, when LLMs still faced significant limitations in context length and inference capabilities. Our study explores techniques that enhance the usability and reliability of LLMs, emphasizing the ongoing importance of well-designed prompts in optimizing task performance. Our findings highlight the critical role of prompt patterns in maximizing LLM's potential, even as their capabilities continue to evolve.
\end{abstract}

\maketitle

Artificial intelligence (AI) has revolutionized software engineering by integrating large language models (LLMs) such as OpenAI's Generative Pre-trained Transformer (GPT) and Codex, which have demonstrated notable capabilities in code generation and developer assistance \cite{brown_language_2020}\cite{chen_evaluating_2021}. Despite the burgeoning application of these approaches, the domain of prompt engineering, i.e., the practice of tailoring inputs to direct LLM behavior toward specific objectives, remains fragmented and lacks comprehensive understanding \cite{srivastava_beyond_2023}.

Software engineering translates human expertise into machine-readable formats, with prompt engineering as a key protocol \cite{zhou_large_2023}. Originating from architecture \cite{a_pattern_language}, patterns have been widely adopted in software engineering
and now extend to generative AI \cite{White2024}. Given the field's variability, abstract patterns offer adaptable solutions, enabling effective responses to unique project demands.

Although the potential of LLMs in software engineering is being explored \cite{zhao_survey_2023}, systematic assessments of how to effectively employ prompt engineering across diverse software engineering contexts are scant. Existing reviews highlight LLM's abilities \cite{hou_large_2023} but lack consolidated knowledge of prompt strategies and their implications for software engineering tasks \cite{guo_evaluating_2023}.

The Prompt Engineering Pattern is a broader framework beyond prompt patterns, encompassing higher-level adjustments that range from tuning the model or its parameters for specific requirements to crafting prompts strategically. This holistic approach includes the following components:

\begin{itemize}
    \item \textbf{Prompt design}: Developing reusable prompt patterns that help users effectively explain their tasks to the LLM.
    \item \textbf{Model tuning}: Fine-tuning the model to suit specific applications better, enhancing its understanding and the relevancy of its output.
    \item \textbf{Parameter adjustment}: Modifying runtime settings such as temperature or top-p to tailor the model's generative behaviors.
\end{itemize}

This comprehensive approach enables a more nuanced interaction with LLMs, allowing for precise and effective solutions that are tailored to complex problems. By defining these patterns, our study aims to encourage more systematic and reliable applications of LLMs in the software engineering industry.

In this article, we conduct a systematic literature review (SLR) for using LLMs in software engineering to answer the following research questions (RQs). We aim to organize our findings into a comprehensive taxonomy that clarifies the nature, roles, and implications of prompt engineering within the software engineering field, moving toward the strategic utilization of LLMs beyond merely superficial applications in software engineering activities.

\begin{itemize}[labelsep=2mm, leftmargin=3mm, align=left, labelwidth=*, itemindent=0cm, itemsep=0.59mm]
\item[\textbf{RQ1.}] \textbf{Which prompt engineering patterns are utilized to enhance the functionalities of LLMs in software engineering tasks?}

\item[\textbf{RQ2.}] \textbf{What are the attributes of prompt engineering patterns?}

\item[\textbf{RQ3.}] \textbf{What are the possible research directions?}
\end{itemize}

This article's primary contribution is its SLR of prompt engineering patterns in software engineering. We propose a taxonomy of these patterns, emphasizing their role in improving the accuracy, efficiency, and scalability of LLM-driven tasks. By analyzing both the state-of-the-art techniques from 2023 and emerging capabilities of LLMs in 2024, we demonstrate that effective prompt patterns are crucial for maximizing the potential of LLMs. This study provides insights to help researchers and practitioners better understand and apply these techniques in the evolving landscape of LLM applications.\footnote{This article substantially extends our preliminary article presented at 2024 COMPSAC \cite{sasaki_systematic_2024}. We have reviewed and added patterns, revised the descriptions, and included a quality-based pattern analysis.}

\section{Related Work}
\label{sec:relatedwork}

In the present study, we build on the insights from several pivotal studies on LLMs and their application in software engineering.

Several reviews of the literature have explored the use of LLMs in software engineering. Hou et al. \cite{hou_large_2023} comprehensively analyzed strategies to optimize LLM performance across software tasks. Zhao et al. \cite{zhao_survey_2023} examined the evolution and challenges of LLM development and applications. Fan et al. \cite{fan_bibliometric_nodate} conducted a bibliometric review of more than 5000 LLM publications, revealing research trends across fields. Although these studies improve understanding of LLMs' capabilities, they do not specifically address prompt engineering in software engineering.

In recent work, White et al. \cite{White2024} bridge this gap by introducing a catalog of reusable prompt patterns that is tailored to software engineering, addressing tasks such as requirement elicitation, design, and code quality improvement. Their structured approach provides a practical framework for leveraging LLMs effectively in these areas.

Our study builds upon these insights by incorporating recent prompt patterns into our taxonomy, addressing the need for updated and practical methodologies. This integration reflects the growing emphasis on systematically applying LLMs in software engineering tasks.

\section{Systematic Literature Review}
\label{sec:review}

We conducted an SLR by using Preferred Reporting Items for Systematic reviews and Meta-Analyses (PRISMA) \cite{prisma_2020}, a widely recognized framework that ensures transparency and reproducibility in research processes. We applied key steps from PRISMA as follows:

\begin{itemize}[labelsep=1mm, leftmargin=1mm, align=left, labelwidth=*, itemindent=0cm, itemsep=0.59mm]
\item \textbf{Eligibility criteria}: Targeted studies on LLM prompts in software engineering.

\item \textbf{Information sources}: Initial searches on IEEE Xplore yielded no relevant studies, prompting an expanded search on Google Scholar. 

\item \textbf{Search strategy and selection}: Using keywords (``prompt engineering'') and (``software engineering'' or ``software development''), we identified 149 studies, later expanded to 158 with additional sources. Screening titles and abstracts reduced these to 37, with 28 selected for in-depth analysis.

\item \textbf{Data collection and synthesis}: Our team conducted data extraction and identified 26 prompt engineering patterns, which we synthesized into a taxonomy.

\end{itemize}

This structured approach ensures clarity and reliability in defining prompt engineering patterns for software engineering tasks.

\section{Taxonomy of Prompt Engineering Patterns}

\begin{figure}[t]
    \centering
    \includegraphics[width=1.04\linewidth]{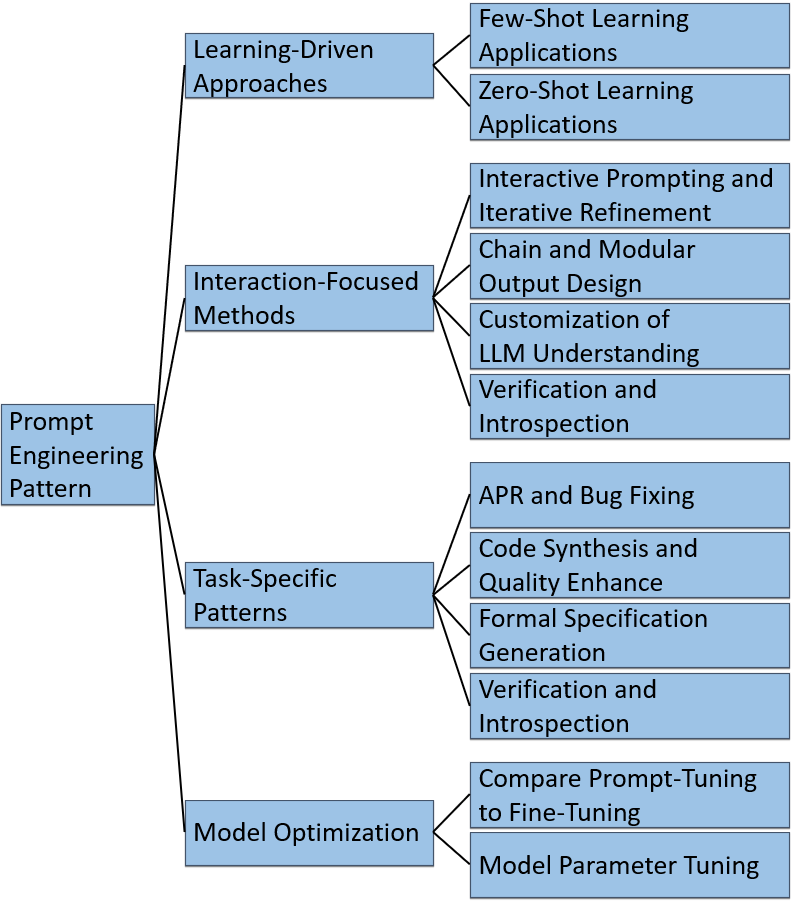}
    \caption{Taxonomy diagram. APR: automatic program repair.}
    \label{fig:taxonomy_diagram}
\end{figure}

To systematically categorize the diverse prompt engineering patterns, we developed a taxonomy, as shown in Table \ref{tab:prompt_pattern}, that addresses RQ1. The taxonomy encapsulates four major categories: learning-driven approaches, interaction-focused methods, task-specific patterns, and model optimization, each with distinct subcategories that are visualized in Figure \ref{fig:taxonomy_diagram}.\footnotetext[2]{The articles for each pattern are listed at https://doi.org/10.5281/zenodo.14281297.\label{patterns_footnnote}}

\begin{table*}[!ht]
    \vspace*{-37pt}
    \caption{
    Categories of prompt engineering pattern.\footref{patterns_footnnote}
    }
    \label{tab:prompt_pattern}
    \scriptsize
    \begin{tabular}{|p{0.03\textwidth}|p{0.095\textwidth}|p{0.095\textwidth}|p{0.155\textwidth}|p{0.255\textwidth}|p{0.265\textwidth}|p{0.140695\textwidth}|}
    \hline
        ID & Category & Subcategory & Pattern Name & Problem & Solution
        \\ \hline
        PA1 & Learning approaches & Few-shot learning applications & Example-based few-shot prompting for code generation %
        & Improving LLMs' code generation efficiency for various tasks without extensive customization or fine-tuning.
        & Employing few-shot learning in LLMs with strategic examples or prompts for efficient, tailored code generation.
        \\  \cline{1-1} \cline{4-6}
        PA2 & ~ & ~ & Abstraction and clue integration for code interpretation %
        & Efficiently retrieve, interpret, and structure SE knowledge while addressing context-specific needs.
        & Design few-shot prompts or abstract code for better code interpretation, quality, and modularity.
        \\  \cline{1-1} \cline{4-6}
        PA3 & ~ & ~ & Example-guided prompting for conflict resolution %
        & Time-consuming manual resolution of textual and semantic merge conflicts in software development.
        & Introduces a merge tool by using k-shot learning for automated merge conflict resolution.
        \\  \cline{1-1} \cline{4-6}
        PA4 & ~ & ~ & Code sample retrieval %
        & Selecting effective prompts for code-related few-shot learning with LLMs.
        & Introduces the automated retrieval technique for selecting relevant code demonstrations for prompts.
        \\  \cline{1-1} \cline{3-6}
        PA5 & ~ & Zero-shot applications & Zero-shot APR and vulnerability repair %
        & Utilizing LLMs for application error explanations and vulnerability patching in zero-shot settings.
        & Strategic zero-shot learning prompts for LLMs to clarify error messages and generate secure software patches.
        \\ \hline
        PB1 & Interaction-focused methods & Interactive prompting and iterative refinement & User-model collaboration refinement %
        & Enhancing interactive dialogue, specification-to-model translation, and AI-assisted coding through improved user-model collaboration.
        & Implement interactive and iterative prompt designs (ReAct) for conversational assistants, object model synthesis, and educational code generation.
        \\ \cline{1-1} \cline{4-6}
        PB2 & ~ & ~ & Prompt translation for clear user intention %
        & Aligning user queries with LLM capabilities for responses that are accurately aligned with user intentions.
        & LLMs use existing prompt patterns to translate user input prompts into clear, system-compatible prompts.
        \\ \cline{1-1} \cline{4-6}
        PB3 & ~ & ~ & Prompt translation for refusal breaker %
        & When LLMs refuse to answer because of their guardrails, users may not know how to rephrase the task. 
        & Add a refusal breaker to explain the reason for the refusal and to suggest alternative wordings.
        \\ \cline{1-1} \cline{4-6}
        PB4 & ~ & ~ & Alternative approach suggestion for task completion %
         & Users often default to familiar methods, potentially missing more efficient or effective alternatives.
         & The LLM suggests alternative methods to achieve the same goal and lists the pros and cons of each method to help users make informed decisions.
        \\ \cline{1-1} \cline{4-6}
        PB5 & ~ & ~ & Context management for improved interaction %
        & LLMs can misinterpret the intention or generate irrelevant responses because of previous unrelated inputs.
        & Enable users to control context by specifying relevant and irrelevant topics, improving the focus of LLM, respectively.
        \\ \cline{1-1} \cline{3-6}
        PB6 & ~ & Chain and modular output design & Gradual execution and output %
        & Enhancing LLMs' performance on complex tasks by improving user control and fostering better collaboration.
        & Break down tasks using chained LLM steps (Chain-of-Thought) to improve answers' quality, accuracy, and transparency.
        \\ \cline{1-1} \cline{4-6}
        PB7 & ~ & ~ & Problem decomposition and integration %
        & Need for a comprehensive method to cover large software requirements.
        & Employ prompt templates for multiclass requirement classification.
        \\ \cline{1-1} \cline{4-6}
        PB8 & ~ & ~ & Infinite output generation %
        & Continuous tasks often require users to enter the same prompt repeatedly.
        & Instruct the LLM to generate several outputs continuously until user stops it.
        \\ \cline{1-1} \cline{3-6}
        PB9 & ~ & Customization of LLM understanding & Domain-specific meta language creation %
        & LLMs may not accurately interpret specialized languages or notations without explicit instructions.
        & Define a meta-language, such as a DSL, and explain its semantics, enabling accurate interpretation and generation.
        \\ \cline{1-1} \cline{3-6}
        PB10 & ~ & Verification and introspection & Fact checklist %
        & LLMs are prone to generating outputs with factual errors or hallucinations, which are often difficult to find.
        & Require LLMs to generate a list of critical facts in their outputs for independent verification.
        \\ \cline{1-1} \cline{4-6}
        PB11 & ~ & ~ & Prompt background reflection %
        & LLMs can provide accurate answers but often lack explanations, making assessing the validity difficult.
        & Instruct the LLM to explain the reasoning and assumptions behind answers to enable users' clearer understanding.
        \\ \hline
        PC1 & Task-specific patterns & APR and bug fixing & Sequential prompting for bug identification and resolution %
        & Resolving software bugs or adapting to new data formats efficiently with traditional APR tools is challenging.
        & Use LLMs to fix or refactor code by providing strategically designed prompts or format-specific examples.
        \\ \cline{1-1} \cline{4-6}
        PC2 & ~ & ~ & Security-focused prompt refinement for hardware debugging %
        & Repairing hardware security bugs in Verilog (an HW-description language) designs by using LLMs.
        & A framework that uses LLMs to generate functionally correct and secure HW code repairs.
        \\ \cline{1-1} \cline{3-6}
        PC3 & ~ & Code synthesis and quality enhance & Leveraging metadata and docs %
        & Code is inherently complex and LLMs struggle to capture its semantics, requiring enriched contextual info.
        & Utilize repository context and tagged identifiers, and generate textual inputs for tools to enhance LLM coding capability.
        \\ \cline{1-1} \cline{4-6}
        PC4 & ~ & ~ & Prompt wording optimization %
        & Enhancing LLM output quality in code summarization through optimized prompt wording.
        & Comparing the output by different prompt wording, like adjusting adverbs, for better quality.
        \\ \cline{1-1} \cline{4-6}
        PC5 & ~ & ~ & Code quality enhancement through detailed prompts and clustering %
        & Exploring the impact of prompt detail level on the syntactic validity, functional correctness, and efficiency of AI-generated code.
        & Systematic evaluation and optimization of prompt details to enhance the quality and practicality of AI-generated code.
        \\ \cline{1-1} \cline{4-6}
        PC6 & ~ & ~ & Comment-driven code quality enhancement %
        & Reducing the reproduction of known bugs in LLM-generated code.
        & Employ strategic comments in prompts to guide LLMs for bug-free coding.
        \\        \cline{1-1} \cline{3-6}
        PC7 & ~ & Formal specification generation & Formal specification generation from natural language %
        & Accurately translating NL specifications into reliable API or object constraint language is difficult.
        & Use structured prompts to generate formal specifications, improving model accuracy and reducing manual effort.
        \\
        \cline{1-1} \cline{3-6}
        PC8 & ~ & Evaluation of code generation security of LLMs & Vulnerability-prone code descriptions %
        & This requires an evaluation dataset to assess the security flaws in code generated by LLMs.
        & Provide a list of NL prompts based on CWEs to benchmark and improve the security of LLM-generated code.
        \\ \hline
        PD1 & Model optimization & Comparison to fine-tuning & Improving code intelligence with prompt tuning %
        & Adapting LLMs to specific code intelligence tasks is difficult because of inconsistent training objectives.
        & Use prompt tuning to adapt models efficiently to code intelligence tasks with limited data.
        \\ \cline{1-1} \cline{3-6}
        PD2 & ~ & Model parameter tuning & Input parameter impact exploration %
        & Leveraging LLMs effectively for diverse and intricate software engineering tasks is complex and inefficient.
        & Design prompt engineering patterns that are tailored for software engineering applications.
        \\ \hline
    \end{tabular}

    \vspace{2pt}
    {\scriptsize API: application programming interface; CWE: Common Weakness Enumeration; DSL: digital subscriber line; HW: hardware; NL: natural language.}
\end{table*}

\subsubsection{Overview of Taxonomy}

Our taxonomy is based on a comprehensive literature review of both peer-reviewed and gray literature, capturing the latest developments in prompt engineering (RQ3). We critically evaluated each study to ensure a balanced representation of the best practices in the field. We categorized the identified prompt patterns into four clusters:

\subsubsection{Learning-Driven Approaches}

Learning-driven approaches encompass few- and zero-shot learning applications that are tailored for specific tasks such as code generation (PA1: Example-based few-shot prompting for code generation) and code interpretation (PA2: Abstraction and clue integration for code interpretation).

\subsubsection{Interaction-Focused Methods}

Interaction-focused methods include patterns that enhance interactive prompting and iterative refinement, such as user-model collaboration refinement (PB1).

\subsubsection{Task-Specific Patterns}

Task-specific patterns cover techniques that are designed for specific software engineering tasks, such as leveraging metadata and documentation (PC3: Leveraging metadata and docs), to improve code generation.

\subsubsection{Model Optimization}

Model optimization involves patterns that are related to model tuning and parameter adjustments to optimize LLM performance.

Each category is linked to a distinct methodological approach toward software engineering activities. For example, in the interaction-focused methods category, patterns such as alternative approach suggestion for task completion (PB4) and gradual execution and output (PB6) emphasize interactive prompting and refinement. In the task-specific patterns category, techniques such as prompt wording optimization (PC4) utilize existing documentation to enhance code generation tasks.

\section{Analysis on RQs}
\label{sec:result}

From our SLR of 158 studies, we selected 28 that applied and evaluated prompt engineering in software engineering, reflecting the growing interest in AI-augmented development practices.

\subsection{RQ1: Which Prompt Engineering Patterns Are Used to Enhance the Functionalities of LLMs in Software Engineering Tasks?}

We identified 26 prompt engineering patterns and classified them into four major categories:

\begin{description}[labelsep=0mm, leftmargin=0mm, align=left, labelwidth=*, itemindent=0cm, itemsep=0.4mm]
\item[1) PA. Learning approach]: Eight articles address the pattern that is used to provide the context necessary for their purpose to LLMs.
The few-shot approach was introduced in many articles, whereas some used zero shot.
\item[2) PB. Interaction-focused methods]: Eight articles discuss methods that are used to enrich the output.
These methods range from submitting relatively simple requests to an LLM, to more extensive task decomposition through mutual communication.
\item[3) PC. Task-specific patterns]: Eight articles discuss the specific patterns of tasks around software engineering.
There were many patterns for code generation or specification, or comment generation from code.
\item[4) PD. Model optimization]: Two articles introduced prompt engineering patterns that are related to model tuning.
The articles in this category introduced prompts to tune the model parameters to specified tasks to elicit the model's potential.
\end{description}

In practical terms, the identified prompt engineering patterns significantly contribute to the refinement of software development methodologies by enabling more precise, efficient, and creative use of LLMs, thereby promising marked improvements in the pace and quality of software engineering tasks.

\textbf{Answer to RQ1: Prompt Engineering Patterns Are Introduced in Many Articles to Enhance the Functionalities of LLMs Across Diverse Software Engineering Activities}

\begin{figure}[t]
    \centering
    \includegraphics[width=1.04\linewidth]{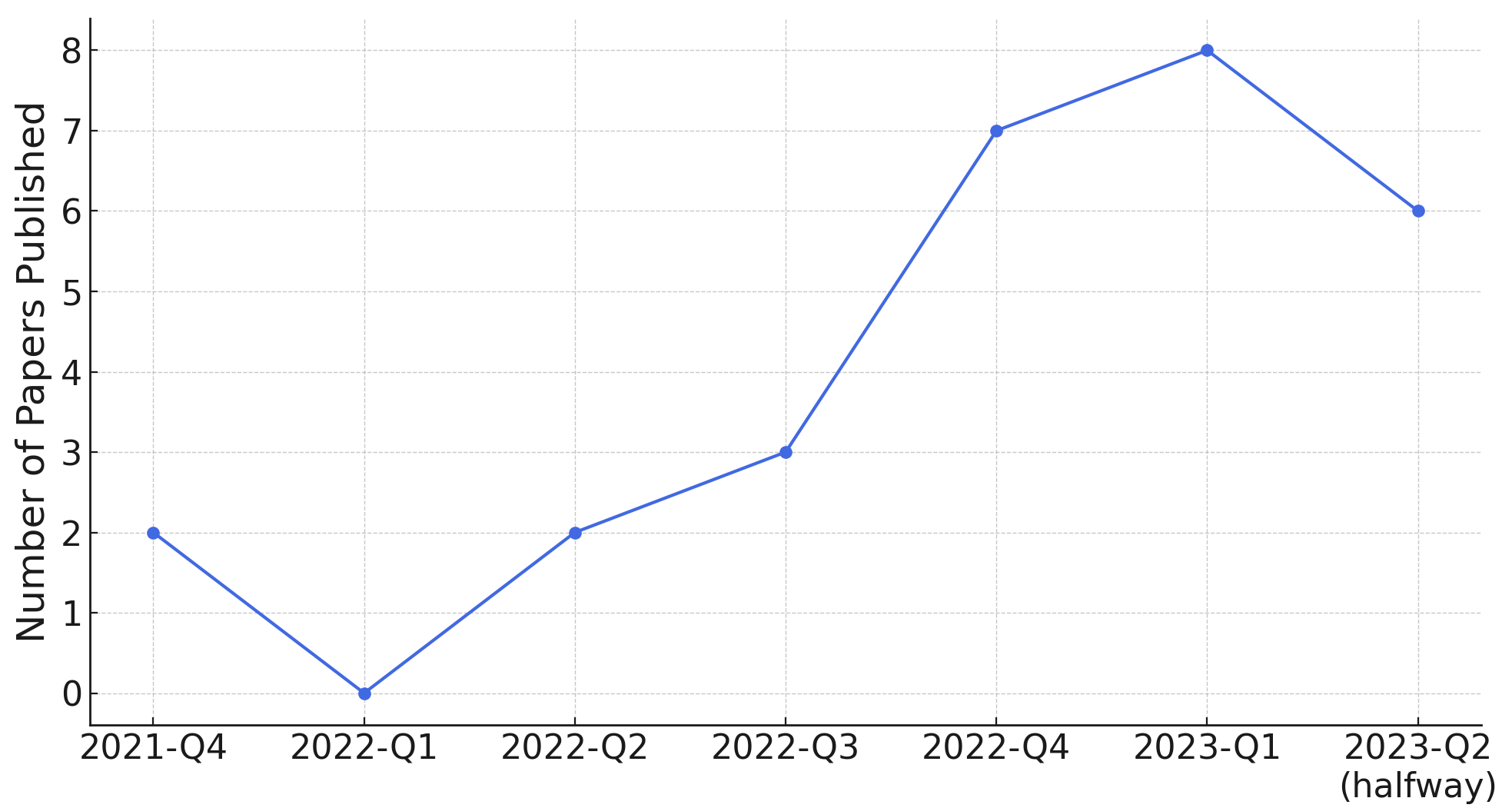}
    \caption{Number of articles published per quarter. Q1: quarter 1; Q2: quarter 2; Q3: quarter 3; Q4: quarter 4.}
    \label{fig:number_of_papers_published_per_quarter_graph}
\end{figure}

\subsection{RQ2: What Are the Attributes of Prompt Engineering Patterns?}

We analyzed the identified patterns regarding publication trends, LLMs used, software development knowledge areas, and the relationship between development knowledge areas and LLMs.

• \textbf{Number of articles published per quarter:} Our analysis reveals a growing research interest in this field, with an evolving approach toward more dynamic and collaborative AI-informed software development processes over time. Figure \ref{fig:number_of_papers_published_per_quarter_graph} illustrates the substantial increase in articles published, from two in 2021 Q4 to eight in 2023 Q1.\newline
• \textbf{LLMs used in articles:} Various LLMs have been explored in the studies, with Codex (or GitHub Copilot) being the most frequently discussed (18 articles), followed by GPT-3 variants (nine articles). Notably, Codex is derived from the GPT-3 architecture and has been fine-tuned specifically for tasks that are related to coding \cite{chen_evaluating_2021}, underscoring the preeminence of the GPT-3 lineage in the field. Other models, such as CodeGen, CodeT5, and PolyCoder, are also referenced, indicating that a broad spectrum of LLMs have been deployed to tackle challenges in software engineering.\newline
• \textbf{Software development knowledge area in articles:} Some articles focus on a specific knowledge area that is defined in the Software Engineering Body of Knowledge \cite{bourque_swebok_2014}; however, most of the articles introduce a more general pattern that is applicable to multiple knowledge areas within software development. The most discussed topic is software construction, followed by testing, maintenance, and design.\newline
• \textbf{Relationship between the development knowledge area and LLMs}: Codex is used predominantly in the construction and testing phases of software development, highlighting its targeted efficacy in code-centric tasks. By contrast, GPT-3 variants are used less frequently. They are applied in various stages of software development, highlighting their broad applicability to tackle a wide range of software engineering challenges beyond simple code generation.\newline
• \textbf{Quality attributes contributed by patterns:} We examined the contributions of these patterns to the software quality attributes that are defined in ISO/IEC 25010:2023 \cite{ISO25010-2023} (see Table \ref{tab:prompt_pattern_and_iso25010}). They enhance predominantly functional suitability and interaction capability, followed by maintainability and flexibility, indicating a strong focus on meeting functional requirements and ensuring user-friendliness.

\begin{table}[t]
    \caption{Patterns and quality attributes from ISO/IEC 25010:2023. FS: functional suitability; PE: performance efficiency; C: compatibility; IC: interaction capability; R: reliability; Se: security; M: maintainability; F: flexibility; SA: safety.}
    \label{tab:prompt_pattern_and_iso25010}
    \begin{tabular}{|c|p{2cm}|c|p{2cm}|}
    \hline
    \textbf{Pattern} & \textbf{QA by pattern} & \textbf{Pattern} & \textbf{QA by pattern} \\ \hline
        PA1 & FS, IC, M, F & PB9 & FS, IC, F \\ \hline
        PA2 & FS, IC, M, F & PB10 & FS, R, Se \\ \hline
        PA3 & FS, PE, C, IC, M, F & PB11 & FS, IC \\ \hline
        PA4 & FS, IC, M, F & PC1 & FS, M, F \\ \hline
        PA5 & FS, Se, M & PC2 & FS, Se, M \\ \hline
        PB1 & FS, IC, M, F & PC3 & FS, IC, M \\ \hline
        PB2 & FS, IC & PC4 & FS, IC \\ \hline
        PB3 & FS, IC, Se & PC5 & FS, PE, M \\ \hline
        PB4 & FS, IC & PC6 & FS, R, M \\ \hline
        PB5 & FS, IC, F & PC7 & FS, IC, M \\ \hline
        PB6 & FS, IC, M & PC8 & FS, Se, M \\ \hline
        PB7 & FS, PE, M & PD1 & FS, PE, M \\ \hline
        PB8 & FS, IC, F & PD2 & FS, PE, F \\ \hline
    \end{tabular}
\end{table}

\begin{table}[t]
    \centering
    \caption{The LLMs used and software development knowledge area covered. con: construction; tes: testing; mai: maintenance; des: design; reqs: requirements.}
    \begin{tabular}{|c|c|c|c|c|c|c|}\hline
    \textbf{LLM}             & \textbf{Con}              & \textbf{Tes}             & \textbf{Mai}              & \textbf{Des}              & \textbf{Reqs}  & \textbf{Total}              \\ \hline
    \textbf{Codex}           & \cellcolor{green!42.86}12 & \cellcolor{green!50}14   & \cellcolor{green!35.71}10 & \cellcolor{green!28.57}8  & \cellcolor{green!14.29}4 & \cellcolor{green!64.29}18  \\ \hline
    \textbf{GPT-3/3.5}       & \cellcolor{green!25}7     & \cellcolor{green!14.29}4 & \cellcolor{green!21.43}6  & \cellcolor{green!21.43}6  & \cellcolor{green!14.29}4 & \cellcolor{green!32.14}9 \\ \hline
    \textbf{CodeGen}         & \cellcolor{green!7.14}2   & \cellcolor{green!7.14}2  & \cellcolor{green!3.57}1   & \cellcolor{green!3.57}1   & \cellcolor{green!3.57}1 & \cellcolor{green!7.14}2  \\ \hline
    \textbf{PolyCoder}       & \cellcolor{green!7.14}2   & \cellcolor{green!3.57}1  & \cellcolor{green!7.14}2   & \cellcolor{green!3.57}1   & \cellcolor{green!0}0  & \cellcolor{green!7.14}2    \\ \hline
    \textbf{Codex}           & \cellcolor{green!7.14}2   & \cellcolor{green!3.57}1  & \cellcolor{green!3.57}1   & \cellcolor{green!3.57}1   & \cellcolor{green!0}0  & \cellcolor{green!7.14}2    \\ \hline
    \textbf{LaMDA}           & \cellcolor{green!3.57}1   & \cellcolor{green!3.57}1  & \cellcolor{green!3.57}1   & \cellcolor{green!3.57}1   & \cellcolor{green!3.57}1  & \cellcolor{green!3.57}1 \\ \hline
    \textbf{PRCBERT}         & \cellcolor{green!3.57}1   & \cellcolor{green!3.57}1  & \cellcolor{green!3.57}1   & \cellcolor{green!3.57}1   & \cellcolor{green!3.57}1 & \cellcolor{green!3.57}1  \\ \hline
    \textbf{GPT-2}           & \cellcolor{green!3.57}1   & \cellcolor{green!0}0     & \cellcolor{green!3.57}1   & \cellcolor{green!3.57}1   & \cellcolor{green!0}0 & \cellcolor{green!3.57}1     \\ \hline
    \textbf{AI21}            & \cellcolor{green!3.57}1   & \cellcolor{green!0}0     & \cellcolor{green!3.57}1   & \cellcolor{green!3.57}1   & \cellcolor{green!0}0 & \cellcolor{green!3.57}1     \\ \hline
    \textbf{GPT-Neo/J/NeoX}  & \cellcolor{green!0}0      & \cellcolor{green!3.57}1  & \cellcolor{green!3.57}1   & \cellcolor{green!0}0      & \cellcolor{green!0}0 & \cellcolor{green!3.57}1     \\ \hline
    \textbf{CodeT5}          & \cellcolor{green!0}0      & \cellcolor{green!3.57}1  & \cellcolor{green!3.57}1   & \cellcolor{green!0}0      & \cellcolor{green!0}0 & \cellcolor{green!3.57}1     \\ \hline
    \textbf{INCODER}         & \cellcolor{green!0}0      & \cellcolor{green!3.57}1  & \cellcolor{green!3.57}1   & \cellcolor{green!0}0      & \cellcolor{green!0}0 & \cellcolor{green!3.57}1     \\ \hline
    \thickhline
    \textbf{Total}         & \cellcolor{green!75}21      & \cellcolor{green!67.85}19  & \cellcolor{green!64.29}18   & \cellcolor{green!57.14}16      & \cellcolor{green!35.71}10 &  x   \\ \hline
    \end{tabular}
\end{table}

\textbf{Answer to RQ2: The Attributes of Prompt Engineering Patterns in Software Engineering Are Characterized by Growing Research Interest, a Notable Focus on the Codex and GPT Models, and Diverse Applications Across Major Software Development Phases and Quality Attributes}

\subsection{RQ3: What Are the Possible Research Directions?}

Future research directions include:

\begin{description}[labelsep=0mm, leftmargin=0mm, align=left, labelwidth=*, itemindent=0cm, itemsep=0.4mm]
    \item[• Underexplored domains]: Expand to areas such as software requirements \cite{marques_using_chatgpt}, leveraging LLMs to streamline elicitation and design processes.
    \item[• Formal analysis]: Develop frameworks for quantitative evaluation of prompt effectiveness in the real world, standardizing metrics across LLMs.
    \item[• Multimodal LLMs]: Leverage LLM advancements to integrate diverse inputs such as diagrams for enhanced task support.
    \item[• Ethics and security]: Address biases and secure practices to build trust in LLM applications.
\end{description}

\textbf{Answer to RQ3: Future Research Should Enhance LLM Capabilities in Underexplored Software Domains, Implement Formal Analysis Frameworks to Validate Implementations Through Real-World Studies, Explore Nontextual Inputs for Multimodal LLMs, and Secure Ethical Standards}

\section{Discussion and Limitation}
\label{sec:discussion}

Our study demonstrates that LLMs are rapidly transforming software engineering, and our taxonomy of prompt engineering patterns aids in their effective utilization. For example, gradual execution and output (PB6) helps developers iteratively refine code, enhancing reliability and time efficiency.

One limitation is that our findings reflect AI models as of mid-2023, when LLMs have shorter context lengths and limited inference capabilities. As LLMs evolve, prompt strategies must adapt to remain effective.

\section{Conclusion and Future Work}
\label{sec:conclusion}
On the basis of a systematic review of recent literature, we developed a practical taxonomy of prompt engineering patterns for software engineering. Our key contributions are summarized as follows:

\begin{description}[labelsep=0mm, leftmargin=0mm, align=left, labelwidth=*, itemindent=0cm, itemsep=0.4mm]
    \item[• Identification of patterns]: We classified prompt engineering patterns that enhance various software development activities, providing a structured approach for developers to use LLMs effectively.

    \item[• Insights on trends]: Our analysis reveals a growing interest in prompt engineering and its potential to address software engineering challenges.

    \item[• Adapting to the evolving LLM landscape]: As LLMs continue to advance, prompt engineering patterns must evolve to effectively leverage new features such as longer context lengths and multimodal inputs.
\end{description}

Future research should focus on expanding the taxonomy into underexplored software domains such as requirement engineering, implementing formal evaluation frameworks for prompt strategies, and utilizing multimodal LLM capabilities. Further investigation into ethical and security considerations is necessary to ensure the responsible deployment of LLMs. Although most of the patterns are abstracted from case studies in the real world, validating these prompt engineering patterns with quantitative measures such as odds ratios could offer additional insight into the patterns' effectiveness. 

Overall, the ongoing integration of LLMs within software engineering requires a dynamic approach to prompt engineering, ensuring that models meet evolving user needs and technological capabilities. Collaborations across academia and industry will be critical to continually optimizing these advancements.

\section{Acknowledgments}
We thank D. Sander for organizing the SLR data. This work was supported in part by the Japan Society for the Promotion of Science (JSPS) KAKENHI Program under Grant 21KK0179 and Grant 23K18470, and in part by the Japan Science and Technology Agency (JST)-Mirai Program under Grant JPMJMI20B8. This article has supplementary downloadable material available at https://doi.org/10.1109/MITP.2024.3525458, provided by the authors.

\def\refname{REFERENCES}
\bibliographystyle{IEEEtran}
\bibliography{prompt_pattern_slr_references}

\begin{IEEEbiography}{Yuya Sasaki} is pursing a master's degree in information science at Waseda University, Tokyo 169-8555, Japan, and is a researcher at ITOCHU Techno-Solutions Corporation. His research interests include the application of artificial intelligence and software engineering. Sasaki received his bachelor's degree in behavioral economics from Waseda University, Japan. Contact him at sasaki.y@ruri.waseda.jp.
\end{IEEEbiography}

\begin{IEEEbiography}{Hironori Washizaki} is a professor and the associate dean at Waseda University, Tokyo 169-8555, Japan. His research interests include systems and software engineering, machine learning software engineering, and ICT education. Washizaki received his Ph.D. degree in information science from Waseda University, Japan. He is the 2025 IEEE Computer Society president. He is a Senior Member of IEEE. Contact him at washizaki@waseda.jp.
\end{IEEEbiography}

\begin{IEEEbiography}{Jialong Li} is a research associate at Waseda University, Tokyo 169-8555, Japan. His research interests include self-adaptive systems and requirement engineering. Li received his M.E. degree in computer science from Waseda University, Japan. He is a Graduate Student Member of IEEE. Contact him at lijialong@fuji.waseda.jp.
\end{IEEEbiography}

\begin{IEEEbiography}{Nobukazu Yoshioka} is the CEO of QAML Inc. and a guest professor at Waseda University, Tokyo 169-8555, Japan, and the Japan Advanced Institute of Science and Technology (JAIST), Ishikawa 923-1292, Japan. His research interests include security and privacy software engineering and software engineering for machine learning-based systems. Yoshioka received his Ph.D. degree in information science from JAIST. Contact him at nobukazu@engineerable.ai.
\end{IEEEbiography}

\begin{IEEEbiography}{Naoyasu Ubayashi} is a professor emeritus at Kyushu University and a professor at Waseda University, Tokyo 169-8555, Japan. His research interests include software engineering, application of artificial intelligence to software engineering, software repository mining, programming language mechanisms, and formal methods/formal verification. Ubayashi received his Ph.D. degree in computer science from the University of Tokyo, Japan. He is an Affiliate Member of IEEE. Contact him at ubayashi@aoni.waseda.jp.
\end{IEEEbiography}

\begin{IEEEbiography}{Yoshiaki Fukazawa} is a professor emeritus at Waseda University, Tokyo 169-8555, Japan, and a professor at the University of Human Environment. His research interests include software engineering, particularly the reuse of object-oriented software. Fukazawa received his Ph.D. degree in electrical engineering from Waseda University, Japan. Contact him at fukazawa@waseda.jp.
\end{IEEEbiography}

\end{document}